\documentclass{PoS}
\pdfoutput=1

\usepackage{amsmath}
\usepackage{amssymb}

\newcommand{\be}{\begin{equation}}
\newcommand{\ee}{\end{equation}}
\newcommand{\K}{{\cal K}}
\DeclareMathOperator{\re}{Re}
\DeclareMathOperator{\myspan}{span}
\DeclareMathOperator{\sign}{sgn}
\DeclareMathOperator{\diag}{diag}

\newcommand{\MathC}{\mathbb{C}}

\title{An iterative method to compute the overlap Dirac operator at nonzero chemical potential}

\ShortTitle{An iterative method to compute the overlap Dirac operator at nonzero chemical potential}

\author{\speaker{Jacques Bloch} and Tilo Wettig\\
        Institute for Theoretical Physics, University of
  Regensburg, 93040 Regensburg, Germany\\
        E-mail: \email{jacques.bloch@physik.uni-regensburg.de}}

\author{Andreas Frommer and Bruno Lang\\
        Department of Mathematics, University of Wuppertal,
  42097 Wuppertal, Germany}

\abstract{
The overlap Dirac operator at nonzero quark chemical potential involves the computation of the 
sign function of a non-Hermitian matrix. In this talk we present an iterative method, first proposed by us in Ref.~\cite{Bloch:2007aw}, which 
allows for an efficient computation of the operator, even on large lattices. 
The starting point is a Krylov subspace approximation, based on the Arnoldi algorithm, for the evaluation of a generic matrix function.
The efficiency of this method is spoiled when the matrix has eigenvalues
close to a function discontinuity. 
To cure this, a small number of critical eigenvectors are added to the Krylov subspace,
and two different deflation schemes are proposed in this augmented subspace.
The ensuing method is then applied to the sign function of the overlap Dirac operator, for two different lattice sizes. The sign function has a discontinuity along the imaginary axis, and
the numerical results show how deflation dramatically improves the efficiency of the method.
}

\FullConference{The XXV International Symposium on Lattice Field Theory\\
		 July 30 - August 4 2007\\
		 Regensburg, Germany}

\begin{document}

\section{The overlap operator and the sign function at nonzero quark chemical potential}
\label{sec:sign}

The overlap Dirac operator \cite{Narayanan:1994gw,Neuberger:1997fp}
provides an exact solution of the Ginsparg-Wilson relation and hence
implements chiral symmetry in lattice QCD even at finite lattice spacing.
At zero quark chemical potential the overlap operator requires the
computation of the sign function of the Hermitian Wilson-Dirac
operator, for which efficient methods have been developed
\cite{Neuberger:1998my,vandenEshof:2002ms}.

To describe QCD at nonzero baryon density (see Ref.~\cite{Stephanov:2007fk} for a review), a quark chemical potential $\mu$ is introduced in the QCD Lagrangian.
The massless overlap Dirac operator at nonzero $\mu$ was defined in Ref.\ \cite{Bloch:2006cd} as
\be 
D_{\text{ov}}(\mu) = 1 + \gamma_5\sign(H_\text{w}(\mu))
\label{Dov}
\ee
with $H_\text{w}(\mu)=\gamma_5 D_\text{w}(\mu)$. $D_\text{w}(\mu)$ is the Wilson-Dirac operator at nonzero chemical potential
\cite{Hasenfratz:1983ba}
\begin{align}
\label{Dw}
[D_\text{w}(\mu)]_{nm} = 
\; & \delta_{n,m} 
   - \kappa \sum_{j=1}^3  (1+\gamma_j) U_{n,j} \delta_{n+\hat j,m} 
   - \kappa \sum_{j=1}^3 (1-\gamma_j)
        U^\dagger_{n-\hat j,j} \delta_{n-\hat j,m} \\
& \phantom{\delta_{n,m}}
   - \kappa  (1+\gamma_4) e^\mu U_{n,4} \delta_{n+\hat 4,m} 
   - \kappa (1-\gamma_4) e^{-\mu}
        U^\dagger_{n-\hat 4,4} \delta_{n-\hat 4,m} \:, \notag
\end{align}
where $\kappa = 1/(8+2m_\text{w})$ with negative Wilson mass $m_\text{w} \in (-2,0)$,
$\gamma_\nu$ with $\nu=1,\ldots,4$ are the Dirac gamma matrices in
Euclidean space, $\gamma_5=\gamma_1\gamma_2\gamma_3\gamma_4$, and
$U_{n,\nu}$ are $SU(3)$ matrices.
The exponential factors $e^{\pm\mu}$ implement the quark chemical potential on the lattice.
For $\mu \ne 0$ the argument $H_\text{w}(\mu)$ of the sign function  in Eq.\eqref{Dov} becomes
non-Hermitian, and one is faced with the problem of defining and computing the sign function of a non-Hermitian matrix.

Consider a given matrix $A$ of dimension $N$ and a generic function $f$. Let $\Gamma$ be a 
collection of closed contours in $\MathC$ such that $f$ is analytic inside
and on $\Gamma$ and such that $\Gamma$ encloses the spectrum of $A$.
Then the function $f( A )$ of the matrix $A$ can be defined by \cite{Dunford}
\be
f(A) = \frac{1}{2\pi i} \oint_\Gamma f(z) (z I - A)^{-1} dz \:,
\label{fcontour}
\ee
where the integral is defined component-wise and $I$ denotes the
identity matrix.
From this definition it is easy to derive a spectral function
definition.
If the matrix $A$ is diagonalizable, i.e., $A=U \Lambda U^{-1}$ with a diagonal
eigenvalue matrix $\Lambda=\diag(\lambda_i)$ and
$U\in\text{Gl}(N,\mathbb C)$, then
\begin{align}
  \label{fA}
  f(A) = U \diag(f(\lambda_i)) U^{-1} \:.
\end{align}
If $A$ cannot be diagonalized, a more general spectral definition can be
derived from Eq.~\eqref{fcontour} using the Jordan decomposition \cite{Golub, Bloch:2007aw}.
Non-Hermitian matrices typically have complex eigenvalues, and applying
Eq.~\eqref{fA} to the sign function in Eq.~\eqref{Dov} requires the
evaluation of the sign of a complex number. 
The sign function needs to satisfy $[\sign(z)]^2=1$ and, for real $x$,
$\sign(x) = \pm 1$ if $x \gtrless 0$.
To satisfy these properties, it has become standard to define
\be
\sign(z) \equiv \frac{z}{\sqrt{z^2}} = \sign(\re(z)) \:,
\label{sgnz}
\ee
where in the last equality the cut of the square root is chosen along the negative real axis. 
Using the definition \eqref{sgnz} in the spectral definition \eqref{fA} and reordering the eigenvalues according to the sign of their real part allows one to write the matrix sign function as
\begin{equation} \label{signA}
\sign(A) = U \left( \begin{array}{cc} +I & \\ & -I \end{array} \right) U^{-1} \:.
\end{equation}
The sign function satisfies $\sign(A)^2 = I$, and a short
calculation \cite{Bloch:2006cd} shows that for this reason the overlap
operator $D_{\text{ov}}(\mu)$ as defined in Eq.~(\ref{Dov}) satisfies
the Ginsparg-Wilson relation.
Moreover, this definition agrees with the result
obtained when deriving Eq.~\eqref{Dov} from the domain-wall fermion
formalism at $\mu\ne0$ \cite{Bloch:2007xi}.

\section{Arnoldi method and function approximation for a non-Hermitian matrix}
\label{NumApp}

A numerical implementation of the sign function using the spectral
definition \eqref{fA} is only possible for small
matrices, as a full diagonalization becomes too expensive as the matrix
grows.
Alternatively, matrix-based iterative algorithms for the
computation of the matrix sign function have been around for many
years, see Ref.~\cite{Higham97} and references therein. These are efficient for medium-sized problems, but are still unaffordable for the very large matrices occurring in typical lattice QCD
simulations.
Therefore, another iterative method is required which approximates 
the vector $y=\sign(A) x$, rather than the full sign matrix itself.  
Such iterative methods are already extensively used for 
Hermitian matrices \cite{Vor88,Drus98}.
Most of these methods are derived from the Lanczos method, which uses  
short recurrences to build an orthonormal basis in a Krylov subspace. 

Krylov subspace methods have also been introduced for non-Hermitian matrices \cite{gallopoulos89parallel,hochbruck}.
The two most widely used methods to compute a basis for the Krylov
subspace are the Arnoldi method and the
two-sided Lanczos method.
In contrast to the Hermitian case, the Arnoldi method requires long
recurrences to construct an orthonormal basis for the Krylov subspace,
while the two-sided Lanczos method uses two short recurrence relations at the cost of losing orthogonality. 
Here we describe a Krylov subspace approximation based on the Arnoldi method to evaluate $f(A)x$ for a generic function of a non-Hermitian matrix. 

We aim to construct an approximation to $f(A)x$ using a polynomial of degree $k-1$ with $k \ll N$. 
For any $k$ there exists a \emph{best} polynomial approximation
$\hat y = P_{k-1}(A) x$ of degree at most $k-1$, which is the orthogonal projection of $f(A)x$ on the Krylov subspace ${\cal K}_{k}(A,x) = \myspan(x, Ax, \ldots, A^{k-1} x)$.
An orthonormal basis $V_k=(v_1,\ldots,v_k)$ for the
Krylov subspace $\K_k(A,x)$ is constructed using the Arnoldi recurrence
\be
A V_k = V_k H_k + \beta_k v_{k+1} e_k^T \:,
\label{Arnoldi}
\ee
where $v_1=x/\beta$, $\beta=|x|$, $H_k$ is an upper Hessenberg matrix, $\beta_k=H_{k+1,k}$, and $e_k$ is the
$k$-th basis vector in $\MathC^{k}$. 
Then $V_k V_k^\dagger$ is a projector on the
Krylov subspace, and the projection
$\hat y$ of $f(A) x$ on $\K_k(A,x)$ can formally be written as
\be
\hat y = V_k V_k^\dagger f(A) x \:.
\label{yproj2}
\ee
However, to compute the projection \eqref{yproj2} one would already have to know the exact result $f(A) x$. 
Therefore, a method is needed to approximate the projected vector
$\hat y$.  
From Eq.~\eqref{Arnoldi} it follows that
\be
H_k = V_k^\dagger A V_k \:,
\label{HVAV}
\ee
which suggests the approximation
\cite{gallopoulos89parallel}
\be
f(H_k) \approx V_k^\dagger f(A) V_k  \:.
\label{fAapprox}
\ee
As $x = \beta V_k e_1$, Eq.\  \eqref{fAapprox} can be substituted in Eq.\ \eqref{yproj2}, finally yielding the approximation
\be
\hat y \approx \beta V_k f(H_k) e_1 \:.
\label{yproj3}
\ee
In this approximation the
computation of $f(A)$ is replaced by that of $f(H_k)$, where $H_k$ is of much smaller size
than $A$. $f(H_k) e_1$ should be evaluated by some suitable numerical method.  

The computation of the matrix sign function using Eq.\ \eqref{yproj3}
converges to the exact solution (see the $m=0$ curve in the left pane
of Fig.~\ref{DeflArn}).  Unfortunately, in the case of the sign
function, the convergence as a function of the size of the Krylov
subspace is very slow if some of the eigenvalues are close to the
function discontinuity along the imaginary axis.  This problem can be
resolved by deflation of these critical eigenvalues.

For Hermitian matrices, it is well known that the computation of the
sign function can be improved by deflating the eigenvalues smallest in
absolute value \cite{vandenEshof:2002ms}.
Assume that $m$ critical eigenvalues $\lambda_i$ of $A$ with orthonormal eigenvectors $u_i$
have been computed. Then
\be
f(A) x = \sum_{i=1}^m f(\lambda_i) ( u_{i}^{\dagger} x )
u_i + f(A) x_\perp \:,
\label{fAxdefl}
\ee
where
$x=x_\parallel + x_\perp$
with $x_\parallel=\sum_{i=1}^m ( u_{i}^{\dagger} x ) u_i$ and
$x_\perp = x - x_\parallel$.
The first term on the right-hand side of Eq.~\eqref{fAxdefl} can be
computed exactly, while the second term can be approximated using 
a Krylov subspace method for $f(A) x_\perp$. Deflation will allow for a much smaller-sized Krylov subspace.

For non-Hermitian matrices the eigenvectors are no longer orthogonal, 
and the simple decomposition into orthogonal subspaces, leading to Eq.\ \eqref{fAxdefl}, no longer holds.
In the next two sections we will 
develop two alternative deflation schemes for the 
non-Hermitian case.

\section{Schur deflation}
\label{Schurdefl}

We construct the subspace $\Omega_m + {\cal K}_k(A,x)$, which is the sum of
the subspace $\Omega_m$ spanned by the right eigenvectors corresponding
to $m$ critical eigenvalues of $A$ and the Krylov subspace ${\cal K}_k(A,x)$.
Assume that $m$ critical eigenvalues and right eigenvectors of $A$ have been computed.
From this, one can construct $m$ Schur vectors $s_i$, which form an orthonormal basis of 
$\Omega_m$, satisfying
\be
A S_m = S_m T_m\:,
\label{pSd}
\ee
where $S_m=(s_1,\ldots, s_m)$ and $T_m$ is an $m \times m$ upper triangular matrix whose diagonal
elements are the eigenvalues corresponding to the Schur
vectors.

We propose a modified Arnoldi method 
to construct an orthogonal basis of the composite subspace
$\Omega_{m} + {\cal K}_k(A,x)$.
That is, each Arnoldi vector is orthogonalized not only
against the previous ones, but also against the Schur vectors
$s_{i}$.
In analogy to \eqref{Arnoldi}, this process can be summarized as
\be
A \begin{pmatrix} S_{m} & V_{k} \end{pmatrix}
=
\begin{pmatrix} S_{m} & V_{k} \end{pmatrix}
\begin{pmatrix} T_{m} & S_{m}^{\dagger} A V_{k} \\
		 0 & H_{k} \end{pmatrix}
+ \beta_k v_{k+1} e_{m+k}^T 
\label{eq:modArnoldi}
\ee
with $v_1=x_\perp/\beta$, where
$x_\perp=( 1 - S_m S_m^\dagger ) x$ is the projection of
$x$ onto the orthogonal complement $\Omega^\perp$ of $\Omega_{m}$ and $\beta=|x_\perp|$.
The Hessenberg matrix
\be
H = \begin{pmatrix} T_m  & S_m^\dagger A V_k \\
0 & H_k \end{pmatrix} 
\label{eq:newH}
\ee
satisfies a relation similar to Eq.~\eqref{HVAV}, namely
$H = Q^\dagger A Q$, 
where the columns of $Q=(S_m \;\, V_k)$ form an orthonormal basis of $\Omega_{m} + {\cal K}_k(A,x)$.
In analogy to Sec.~\ref{NumApp} we construct the approximation
\be
f(A) x \approx Q f(H) Q^\dagger x \:.
\label{yproj4}
\ee
Because of the block structure \eqref{eq:newH} of $H$, the matrix $f(H)$ can be written as 
\be
f(H) =
\begin{pmatrix}
f(T_m) & Y \\
0 & f(H_k)
\end{pmatrix} \:,
\label{eq:f_H}
\ee
where $Y$ reflects the coupling between both subspaces
and satisfies the Sylvester equation
\be
T_m Y - Y H_k = f(T_m) X - X f(H_k) 
\label{SylvEq}
\ee
with $X=S_m^\dagger A V_k$.
Combining \eqref{yproj4} and \eqref{eq:f_H}, we obtain
\begin{eqnarray}
  f( A ) x
& \approx &
 S_{m} f( T_{m} ) S_{m}^{\dagger} \: x
	+ \begin{pmatrix}
		 S_{m} & V_{k}
	  \end{pmatrix}
	  \begin{pmatrix}
		 Y \\
		 f( H_{k} )
	  \end{pmatrix}
	  \beta e_{1}
\: .
  \label{eq:formula}
\end{eqnarray}
$f(T_m)$ and $f(H_k)e_1$ are computed with some suitable numerical method, 
and the mixed triangular/Hessenberg Sylvester equation \eqref{SylvEq} for $Y$ can be solved efficiently with the method of Ref.~\cite{Bloch:2007aw}.

\section{LR-deflation}
\label{LRdefl}

An alternative deflation in the same composite subspace $\Omega_m +{\cal K}_k(A,x)$ can be constructed using both the left and right eigenvectors corresponding to the critical eigenvalues. 
Assume that $m$ critical eigenvalues of $A$ and their corresponding left and right eigenvectors have been computed,
\begin{align}
A R_m = R_m \Lambda_m \: , \qquad
L_m^\dagger A = \Lambda_m L_m^\dagger\:,
\label{LRev}
\end{align}
where $\Lambda_m$ is the diagonal matrix of critical eigenvalues, and
$R_m=(r_1,\ldots,r_m)$ and 
$L_m=(\ell_1,\ldots,\ell_m)$ are the matrices with the corresponding right and left eigenvectors, respectively.
The left and right eigenvectors corresponding to different eigenvalues are orthogonal. If  the eigenvectors are normalized such that $\ell_i^\dagger r_i=1$, then $L_m^\dagger R_m = I_m$, and $R_m L_m^\dagger$ is an oblique projector on the subspace $\Omega_m$ spanned by the right eigenvectors. 
Let us now decompose $x$ as
$x = x_{\parallel} + x_{\ominus}$, 
where $x_{\parallel} = R_m L_m^\dagger x$ and $x_{\ominus} = x-x_{\parallel}$. Then
\be
f(A) x =  R_m f(\Lambda_m) L_m^\dagger x  + f(A) x_{\ominus} \:.
\label{fAxLR}
\ee
The first term on the right-hand side, which follows from Eq.~\eqref{LRev}, can be evaluated exactly,
while the second term can be approximated by applying the Arnoldi method described in Sec.~\ref{NumApp} to $x_{\ominus}$. An orthonormal basis $V_k$ is constructed in the Krylov subspace ${\cal K}_k(A,x_{\ominus})$ using the Arnoldi recurrence \eqref{Arnoldi}, with $v_1=x_{\ominus}/\beta$ and $\beta=|x_{\ominus}|$. 
Successive operations of $A$ on $x_{\ominus}$ will yield no contributions along 
the $m$ critical eigendirections, hence $\K_k(A,x_\ominus)$ does not mix with $\Omega_m$.
Applying the Arnoldi approximation \eqref{yproj3} to Eq.~\eqref{fAxLR} yields  the final approximation
\be
f(A) x \approx  R_m f(\Lambda_m) L_m^\dagger x  + \beta V_k f(H_k) e_1 \:.
\label{fAxLRArn}
\ee
Again, the first column of $f(H_k)$ will be computed with some suitable numerical method.

\section{Results}
\label{Results}

\begin{figure}
\centering
\includegraphics[width=75mm]{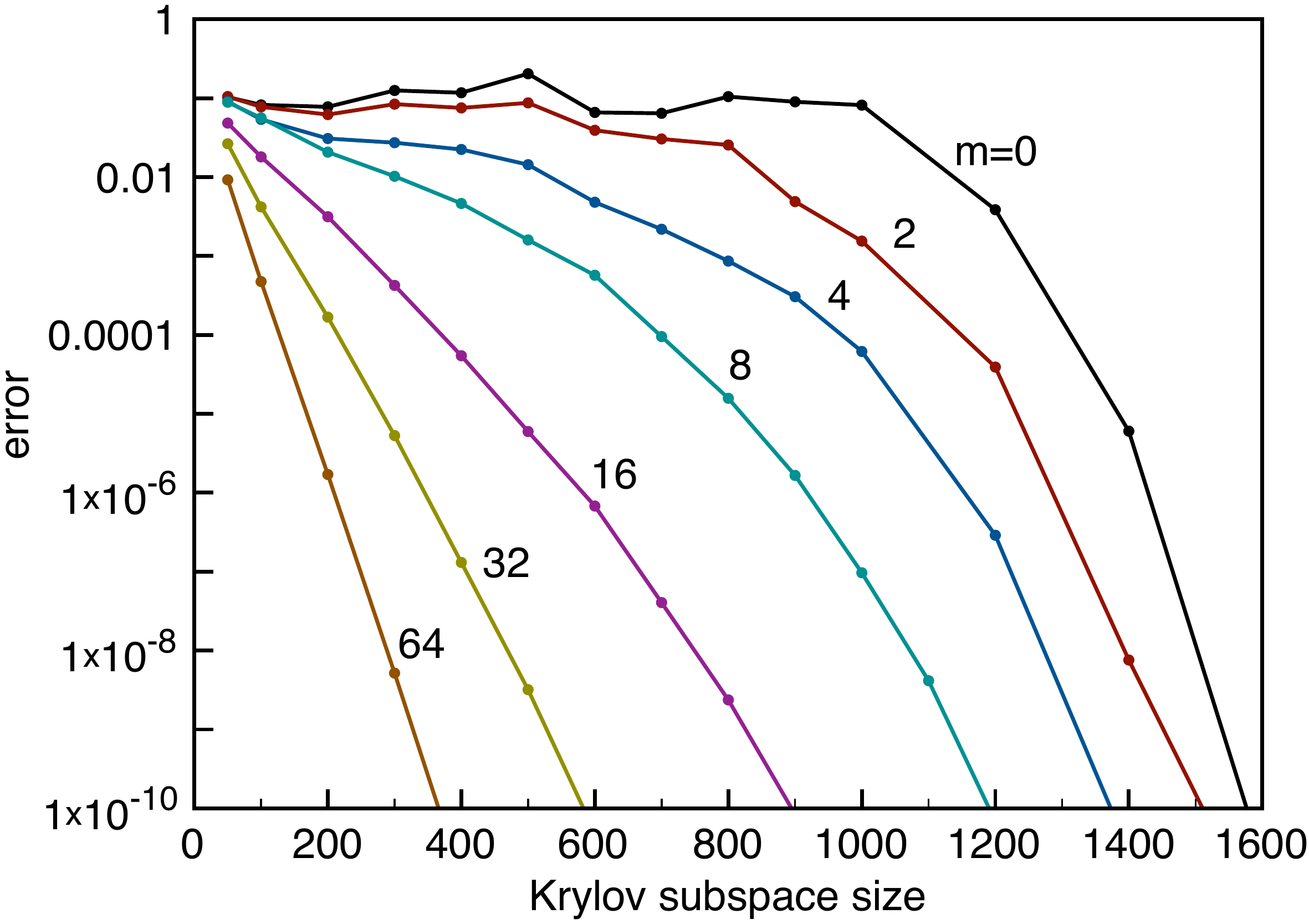}
\includegraphics[width=75mm]{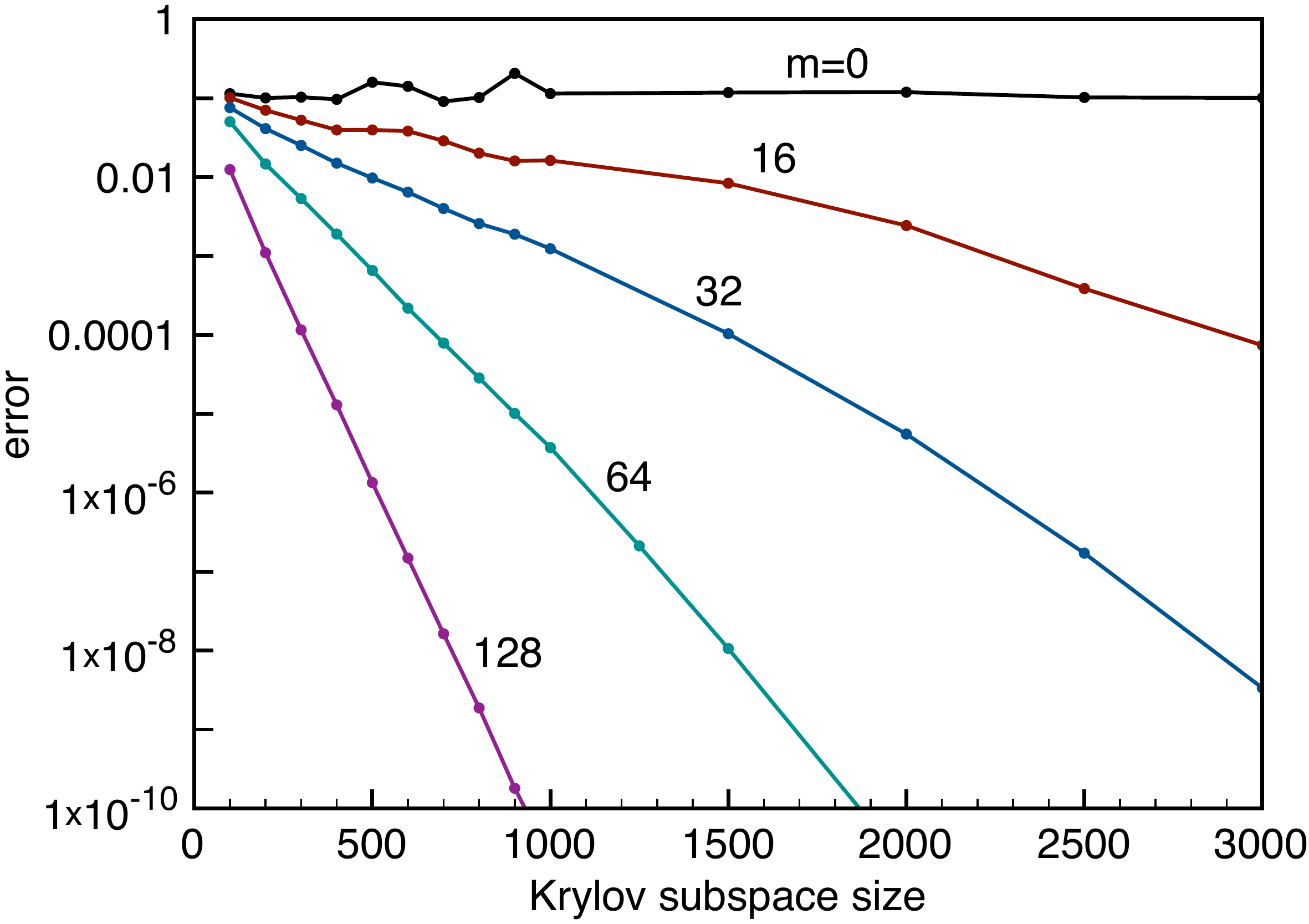}
\caption{\label{DeflArn}Accuracy of approximation \protect\eqref{fAxLRArn} for
   $y=\sign(H_\text{w}(\mu)) x$ 
   with $x=(1,1,\ldots,1)$ at $\mu=0.3$, for a $4^4$ lattice (left) and a
    $6^4$ lattice (right). 
   The relative error $\epsilon=\| \tilde y - y \| / \| y \|
   $ is shown as a function of the Krylov subspace size $k$ for various
   numbers of deflated eigenvalues $m$ using the LR-deflation.
   }
\end{figure}

We used the methods described above to compute the sign function occurring in the overlap Dirac operator \eqref{Dov} for a $4^4$ and a $6^4$ lattice gauge configuration. 
Deflation of critical eigenvalues is essential because $H_\text{w}(\mu)$ has eigenvalues close to the imaginary axis. 
In practice, these eigenvectors need to be computed to high accuracy, as this will limit the overall accuracy of the function approximation. This was done with ARPACK \cite{arpack}. The modified Arnoldi method was implemented in C++ using the optimized ATLAS BLAS library \cite{atlas-hp}. 
The convergence of the method is illustrated in Fig.~\ref{DeflArn},
where the accuracy of the approximation is shown as a function of the
Krylov subspace size.
The various curves correspond to different numbers of deflated
eigenvalues,  using the LR-deflation scheme.
Without deflation ($m=0$) the need for large Krylov subspaces would make the method unusable. 
Clearly, deflation highly improves the efficiency of the numerical method:
as more eigenvalues are deflated, smaller Krylov subspaces 
are sufficient to achieve a given accuracy. 
Furthermore, the deflation efficiency seems to grow with increasing lattice volume.
Indeed, although the matrix size $N$ for the $6^4$ lattice is more than 5 times larger than in the $4^4$ case, the Krylov subspace only has to be expanded by a factor of 1.2 to achieve a given accuracy of $10^{-8}$ (for $m \approx 0.008 N$).
It is also interesting to note that the modified Arnoldi approximation \eqref{fAxLRArn}  for $f(A) x$ is very close to the \emph{best} approximation in the composite subspace, which is given by the orthogonal projection of $f(A) x$ on $\Omega_m + {\cal K}_k(A,x)$, as was checked numerically.

The results for the Schur deflation are not shown here, but are very similar to those for the LR-deflation. 
The Schur deflation is slightly less accurate, and requires more CPU time per evaluation, mainly because of the additional orthogonalization of the Arnoldi vectors with respect to the Schur vectors. 
However, the time taken by its initialization phase is halved, as it
only requires the computation of the right eigenvectors, and the
best choice of deflation scheme will depend on the number of vectors $x$ for
which $\sign(H_\text{w}) x$ needs to be computed.  If one needs to apply
both $\sign(H_\text{w})$ and its adjoint, then, obviously, the LR-deflation will
be the better choice.
A more detailed discussion of both deflation schemes can be found in Ref.~\cite{Bloch:2007aw}

\section{Conclusion}

In this talk we presented an algorithm to approximate the action
of a function of a non-Hermitian matrix on an arbitrary vector, when
some of the eigenvalues of the matrix lie in a region of the complex
plane close to a discontinuity of the function.
The method approximates the solution vector in a composite subspace consisting of
a Krylov subspace augmented by the eigenvectors corresponding 
to a small number of critical eigenvalues.
Two deflation variants were presented based on different subspace decompositions: the Schur deflation uses two coupled orthogonal  subspaces, while the LR-deflation uses two decoupled but non-orthogonal subspaces.
Deflation explicitly takes into account the contribution of the
critical eigenvalues. This allows for smaller-sized
Krylov subspaces, which is crucial for the efficiency of the method.
The method was applied to
the overlap Dirac operator of lattice QCD at nonzero chemical potential, 
where the importance of deflation was clearly demonstrated.

\section*{Acknowledgments}
This work was supported in part by DFG grants FOR465-WE2332/4-2 and Fr755/15-1.

\bibliographystyle{h-elsevier3}
\bibliography{biblio}

\end{document}